\documentclass[12pt]{iopart}
\usepackage{iopams} 

\usepackage{times}

\DeclareFontFamily{OT1}{times}{}
\DeclareFontShape {OT1}{times}{m }{n }{ <-> ptmr }{}
\DeclareFontShape {OT1}{times}{bx}{n }{ <-> ptmb }{}
\DeclareFontShape {OT1}{times}{m }{it}{ <-> ptmri}{}
\DeclareFontShape {OT1}{times}{bx}{it}{ <-> ptmbi}{}

\newcommand{\DEF}{:=}   
\newcommand{\eqref}{\eref}
\newcommand{\text}{\mbox}

\newcommand{\HU} {\mathrm{H}}         
\newcommand{\DUP}{\delta}             
\newcommand{\UPS}{\Upsilon}           
\newcommand{\BRA}{\langle\kern -.2em\langle} 
\newcommand{\KET}{\rangle\kern -.2em\rangle} 
\newcommand{ \iint}{\int\kern -.8em\int}
\newcommand{\iiint}{\int\kern -.8em\int\kern -.8em\int}
\def\Mt{\mathrm{M}}       
\def\a{\underline{a}}         
\def\lima{\lim_{       a \rightarrow 0}}
\newcommand{\rmD}{\mathrm{D}}         
\newcommand{\ASS}{\asymp}             
\newcommand{\OOO}{\mathrm{O}}         
\renewcommand{\theequation}{\thesection.\arabic{equation}}

\begin{document}


\title[Colombeau generalized-functions in classical electrodynamics]
      {A concise introduction to Colombeau generalized functions
       and their applications in classical electrodynamics}

\author{Andre Gsponer}

\address{Independent Scientific Research Institute,
         Oxford, OX4 4YS, United Kingdom}

\begin{abstract}

The objective of this introduction to Colombeau algebras of generalized-functions (in which distributions can be freely multiplied) is to explain in elementary terms the essential concepts necessary for their application to basic non-linear problems in classical physics.

Examples are given in hydrodynamics and electrodynamics.  The problem of the self-energy of a point electric charge is worked out in detail: The Coulomb potential and field are defined as Colombeau generalized-functions, and integrals of nonlinear expressions corresponding to products of distributions (such as the square of the Coulomb field and the square of the delta-function) are calculated.

Finally, the methods introduced in Eur.~J.~Phys.~{\bf 28} (2007) 267, 1021, and 1241, to deal with point-like singularities in classical electrodynamics are confirmed.  

\end{abstract}

\section{Introduction} 
\label{int:0} \setcounter{equation}{0}

The theory of distributions, invented by Laurent Schwartz nearly 60 years ago,  provided a simple and rigorous calculus unifying a great variety of previously ill-defined mathematical techniques used in physics and engineering.  Improper functions such as Heaviside's step function $\HU(x)$ and its derivative, Dirac's delta function $\delta(x)$, were replaced by linear functionals called `distributions,' which like $\mathcal{C}^\infty$ functions can be differentiate any number of times \cite{CHOQU1982-,SCHUC1991-}.

   Distributions, however, cannot in general be multiplied. In mathematical language the set $\mathcal{D}'$ of all distributions is a vector space rather than an algebra.  But products of distributions arise naturally in many areas of science and engineering, most prominently in electrodynamics and particle physics as self-interaction terms of the type $(1/x^2)^2$ and $\delta^2(x)$, and in hydrodynamics as products like $\HU(x)\delta(x)$ of shock waves and their derivatives, as well as in mathematics as solutions of partial differential equations.  Many proposals have therefore been made to define an algebra of generalized functions $\mathcal{G}$ such that $\mathcal{D}' \subset \mathcal{G}$.

    The difficulty of this task is not only mathematical (Schwartz's theory of distribution is highly abstract) but also conceptual since many generalizations are possible.  In this sense the algebras defined by Jean-Fran{\c{c}ois Colombeau \cite{COLOM1984-,COLOM1985-,COLOM1990-} have essentially optimal properties for a wide range of applications \cite{SCHME1990-,COLOM1992-,HORMA1998-,STEIN2006-}, which combined with the fact that they provide what is possibly the most simple and natural generalization of the space of distributions, have made them very popular (Ref.~\cite{GROSS2001-} and numerous references therein).

  Trying to make these developments accessible to as wide an audience as possible is also a difficult task: This paper is therefore a rather specialized article, at the crossroad of modern mathematics and physics, which would only be accessible to advanced undergraduates and graduate students (and their teachers) in mathematical and theoretical physics. 

  The ambition of this paper is to give a straightforward introduction to Colombeau generalized functions and to their applications by emphasizing the underlying concepts rather than the mathematical details of the theory, in order to make it accessible to most physicists and engineers.  The idea is that working with Colombeau functions is like working with real numbers:  There is no need to know how transcendental numbers are embedded in the set of real numbers to use them.

   In particular, we do not introduce Colombeau's algebras by starting with their abstract definition as quotient spaces $\mathcal{G} = \mathcal{E}_\Mt/\mathcal{N}$, where $\mathcal{E}_\Mt$ is a space of `moderate' (or `multipliable') functions, and $\mathcal{N} \subset \mathcal{E}_\Mt$ an ideal of `negligible' functions such that $\mathcal{E}_\Mt/\mathcal{N}$ becomes a superset of $\mathcal{D}'$.  We prefer to emphasize the fundamental idea behind their construction, which is that the smooth functions (i.e., the $\mathcal{C}^\infty$ functions ---  which are indefinitely continuously differentiable) must be a faithful subalgebra of $\mathcal{G}$.  Moreover, to be self-consistent, we begin by recalling in Sec.\,\ref{def:0} some basic definitions, and in Sec.\,\ref{sch:0} the fundamentals of Schwartz distributions.  In the same spirit we summarize in Sec.\,\ref{mul:0}, that is before defining Colombeau's algebra in Sec.\,\ref{col:0}, the reasons why the product of distributions is inconsistent, and how regularization can be used to circumvent this problem.

   We then consider two sets of applications.  First, in Sec.\,\ref{hyd:0}, to hydrodynamics, the second major topic after quantum field theory to which Colombeau originally applied his theory \cite{COLOM1992-}, which enables to illustrate its power in numerical engineering and as a general method for solving nonlinear partial differential equations.  Second, in Secs.\,\ref{poi:0} to \ref{sen:0}, to classical electrodynamics, which enables to show how the introduction of point charges using the Colombeau formalism leads to the possibility of calculating divergent quantities such as the self-energy --- which are quadratic in the fields, and therefore meaningless in Schwartz distribution theory.  Finally, in Sec.\,\ref{lin:0}, we use the formalism of generalized functions to confirm the simplified methods introduced in References \cite{GSPON2004D} and  \cite{GSPON2007C} to deal with point-like singularities in linear problems of classical electrodynamics.

\section{Definitions and notations} 
\label{def:0} \setcounter{equation}{0}

This section is put here rather than in appendix because it recalls definitions and notions which give an opportunity to anticipate important concepts that will be developed in the text, and because the subject of generalized functions makes it essential to be particularly careful with notations, and with the belonging of all objects to the proper sets.

\begin{itemize}

\item[$\quad \Omega$] An open set of $\mathbb{R}^n$.  In this paper $\Omega$ is any interval $]a,b[ \subset \mathbb{R}$ such that $a<0<b$, unless otherwise specified. 

\item[$\mathbb{N}_0$] The set $\{ 0, \mathbb{N}\}$, i.e., the natural numbers and zero.

\item[$\mathcal{C}^\infty$] The algebra of continuous functions on $\Omega$ which are smooth, i.e., that are indefinitely continuously differentiable, and which have compact support.

\item[$\mathcal{C}^m$] The space of continuous functions on $\Omega$ which are $m$-times continuously differentiable and which have compact support.

\item[$\mathcal{C}$] The space $\mathcal{C}=\mathcal{C}^0$ of continuous functions on $\Omega$ with compact support.  Such functions may have points at which the left and right derivatives are different.

\item[$\mathcal{C}_p$] The space of piecewise continuous functions on $\Omega$ with compact support, i.e., continuous except on a discrete set, on which they have left and right limits, and where differentiation leads to $\delta$-functions.

\item[$\mathcal{D}$] The space of functions $\mathcal{C}^\infty(\Omega)$ equipped with an inductive limit topology suitable to define Schwartz distributions. The functions in this space are used to construct the weakly converging sequencies defining distributions, as well as to provide so-called `test functions,' denoted in this paper by $T$, on which the distributions are evaluated.

\item[$\mathcal{D}'$] The space of Schwartz distributions, i.e., the set of all linear continuous functions (linear forms) on $\mathcal{D}(\Omega)$, that is, the topological dual of $\mathcal{D}(\Omega)$.  

\item[$\mathcal{G}$] The algebra of Colombeau generalized functions, or $\mathcal{G}$-functions, on $\Omega$.

\item[~] The Colombeau algebra $\mathcal{G}$, as well as three function spaces: $\mathcal{E}, \mathcal{E}_{\text{M}}$, and $\mathcal{N}$, will be defined in Sec.\,\ref{col:0}.

\end{itemize}

  In the sequel we will generally keep $\Omega$ implicit, and tacitly assume that all functions are zero outside of $\Omega$.  Finally, we will use the term `sequence' for expressions such as $\lim_{\epsilon \rightarrow 0} F_\epsilon$ even though mathematicians reserve this term to mappings $n \mapsto F_n$, so that $\epsilon \rightarrow 0$ corresponds to $1/n \rightarrow 0$ as $n \rightarrow \infty$.

\section{Schwartz distributions} 
\label{sch:0} \setcounter{equation}{0}

  The space $\mathcal{D}'$ of Schwartz distributions contains, besides the ordinary (i.e., $\mathcal{C}$, $\mathcal{C}^m$ and $\mathcal{C}^\infty$) functions, generalized functions corresponding to discontinuous functions and unbounded functions.  While these functions cannot be differentiated in the classical sense, they can be indefinitely differentiated in the sense of distributions.  That is, if $D \in \mathcal{D}'$ is any distribution, its derivatives in the `distributional sense' are such that, $\forall T \in \mathcal{D}$ and $\rmD^n = \partial^n/\partial x^n$,
\begin{eqnarray} 
\label{sch:1}
  \int_\Omega dx ~(\rmD^n D \bigr )(x)~T(x)  \DEF (-1)^n
  \int_\Omega dx ~D(x) ~(\rmD^n T \bigr )(x).     
\end{eqnarray}

     It is not possible to represented non-trivial distributions, such as Dirac's $\delta$-function, by simple algebraic formulas or even by ordinary limiting processes.  They can however be represented by sequences of smooth functions 
\begin{eqnarray} 
\label{sch:2}
  D(x) \DEF \lim_{\epsilon \rightarrow 0} D_\epsilon(x),
     ~~~ \text{where} ~~~ D_\epsilon(x) \in \mathcal{D},
\end{eqnarray}
for which ordinary pointwise convergence is not required.  Instead, what is required is `weak convergence' for the scalar product of $D(x)$ with any test functions $T(x) \in \mathcal{D}$, i.e., the existence of the limit
\begin{eqnarray} \label{sch:3}
   \forall T \in \mathcal{D}, ~~~ ~~~\BRA D | T \KET \DEF 
   \lim_{\epsilon \rightarrow 0}\int_\Omega D_\epsilon(x) T(x)~dx
   \in \mathbb{R}.
\end{eqnarray}
The meaning of operating `in the sense of distributions' is then that all operations on distributions are actually performed on $D_\epsilon(x)$, while $D(x)$ can be seen as a convenient symbol to designate a given distribution.

    Eq.~\eqref{sch:3} shows that distributions can be interpreted as linear functionals $D(T) = \BRA D | T \KET$ defined by their effect on test functions.   Moreover, since many different sequencies may converge weakly to the same limit, each distribution corresponds to an equivalence classes of such sequencies, which all together form the Schwartz distribution space $\mathcal{D}'$:
\begin{quote}\label{defi:1}{\bf Definition 1}
{\it Two distributions $D$ and $E \in \mathcal{D}'$, of respective representatives  $D_{\epsilon}$ and $E_{\epsilon}$, are said to be equal (or equivalent), and one write $D=E$, iff}
\begin{eqnarray} \label{sch:4}
 \forall T \in \mathcal{D}, ~~~ ~~~ \lim_{\epsilon \rightarrow 0} 
           \int_{\Omega} \Bigl(D_{\epsilon}(x) - 
                         E_{\epsilon}(x)\Bigr)T(x) ~ dx = 0.
\end{eqnarray}
\end{quote}
For instance, Dirac's $\delta$-function is defined by the property
\begin{eqnarray} \label{sch:5}
   \BRA \delta | T \KET = \int_\Omega \delta(x) T(x)~dx = T(0),
\end{eqnarray}
so that all sequences which have this property form an equivalence class corresponding to Dirac's $\delta$-function distribution, conventionally denoted by the symbol `$\delta(x)$.'  Two examples of such sequences are
\begin{eqnarray} \label{sch:6}
   \delta_\epsilon(x) = \frac{1}{\pi\epsilon}
                        \frac{\epsilon^2}{\epsilon^2 +x^2},
   ~~~ ~~~ \text{and} ~~~ ~~~ 
   \delta_\epsilon(x) = \frac{1}{\sqrt\pi\epsilon}
                        \exp\bigl(-\frac{x^2}{\epsilon^2}\bigr).
\end{eqnarray}
More generally, any normalizable $\mathcal{C}^\infty$ function $\rho(y)$  with compact support can be used to define $\delta$-sequencies, i.e.,
\begin{eqnarray} \label{sch:7}
   \forall \rho \in \mathcal{C}^\infty_0, ~~~ ~~~  \int_\Omega \rho(y) ~ dy = 1 
   ~~~ ~~~ ~~ \Rightarrow ~~~ ~~~ ~~
   \delta_\epsilon(x) \DEF \frac{1}{\epsilon} \rho\bigl(\frac{x}{\epsilon}\bigr).
\end{eqnarray}

   The requirement that both $\delta_\epsilon$ and $T$ are $\mathcal{C}^\infty_0(\Omega)$ functions and the definition \eqref{sch:1} enable to derive at once a number of useful properties.  For instance, the equations
\begin{eqnarray} \label{sch:8}
   \BRA x \delta(x) | T(x) \KET = 0,
   ~~~ ~~~ \text{and} ~~~ ~~~ 
   \BRA \frac{1}{x}\delta(x) | T(x) \KET = -T'(0),
\end{eqnarray}
which are often symbolically written `$x\delta(x) = 0$,' and  `$x^{-1}\delta(x) = - \delta'(x)$' or `$x\delta'(x) = - \delta(x)$' are the fundamental formulas of calculus with distributions.\footnote{As will be seen, the Colombeau formalism provides a non-ambiguous notation for these formulas, i.e., $x\delta(x) \ASS 0$, and $x^{-1}\delta(x) \ASS - \delta'(x)$ or $x\delta'(x) \ASS - \delta(x)$.}

  In summary distributions are not functions in the usual sense but equivalence classes of weakly convergent sequencies of smooth functions.  All operations on distributions are therefore made on these sequencies, which are thus added, differentiated, etc., according to the operation in question.  It is remarkable that distributions enjoy essentially all properties of $\mathcal{C}^\infty$ functions, including multiplication by a $\mathcal{C}^\infty$ function, with a few exceptions such as the impossibility to multiply two distributions in the general case, as was demonstrated by Schwartz in his impossibility theorem of 1954 \cite[p.\,8]{COLOM1992-}, \cite[p.\,6]{GROSS2001-}.

  Moreover, Schwartz distributions have a very precise relation to continuous functions, which can be spelled in the form of the theorem:
\begin{quote}{\bf Theorem 1 (Schwartz local structure theorem)}\label{theo:1}
{\it Any distribution is locally a partial derivative of a continuous function \emph{\cite[p.\,6]{COLOM1992-}}.}
\end{quote} 
Differentiation induces therefore the following remarkable cascade of relationships: {\it continuously differentiable functions $\rightarrow$ continuous functions $\rightarrow$  distributions}.  This gives a unique position to Schwartz distributions because they constitute the smallest space in which all continuous functions can be differentiated any number of times.  For this reason it is best to reserve the term `distribution' to them, and to use the expression `generalized function' for any of their generalizations.

\section{Multiplication and regularization of distributions} 
\label{mul:0} \setcounter{equation}{0}

  There are two kinds of problems with the multiplication of distributions:  (i) The product of two distributions is, in general, not defined.  For example, the square of Dirac's $\delta$-function is not a weakly converging sequence, as can easily be verified by squaring either of the sequencies in Eq.~\eqref{sch:6} and trying to evaluate them in a scalar product with any test function.  (ii) Differentiation is inconsistent with multiplication because the Leibniz rule, or even associativity, can fail under various circumstances.  For example, while Dirac's $\delta$-function is related to Heaviside's step function through differentiation as $\delta(x) = \HU'(x)$, the algebraic identity $\HU^2(x)=\HU(x)$ leads to inconsistencies.  Indeed,
\begin{eqnarray} \label{mul:1}
         \HU^2 = \HU       ~~ \Rightarrow ~~
    2\HU\delta =    \delta ~~ \Rightarrow ~~
  2\HU^2\delta = \HU\delta ~~ \Rightarrow ~~
    2\HU\delta = \HU\delta ~~ !
\end{eqnarray}

   Over the years many methods for solving these problems have been proposed.  One of the simplest and most effective is `regularization,' which consists of modifying the functions to be multiplied or differentiated in such a way that they become more regular (i.e., continuous, differentiable, finite, etc.).  All operations are then done with the regularized functions until the end of the calculation, and the final result is obtained by the inverse process which returns the function from its regularization.

  A particularly convenient regularization technique is based on the convolution product.  For instance, if $f(x)$ is any function on $\mathbb{R}$, its regularization $f_\epsilon(x)$ is
\begin{eqnarray} \label{mul:2}
      f_\epsilon(x) = (f * \rho_\epsilon)(x)
     \DEF \int_{\Omega} f(x-y) \rho_\epsilon(y) ~dy.
\end{eqnarray}
Here $\rho_\epsilon(x)$ is a smoothing kernel (also called regularizer or `mollifier') which in its simplest form is a $\delta$-sequence as defined in Eq.~\eqref{sch:7}, and $\epsilon \in ]0,1[$ is the (so called) regularization parameter. Consequently, in the limit $\epsilon \rightarrow 0$, the mollifier becomes equal to the $\delta$-function, which by Eq.~\eqref{sch:5} acts as the unit element in the convolution product, i.e., $f*\delta=f$.  Thus, when $\epsilon \neq 0$ the regularization is such that $f$ is `mollified' by the convolution, while $f$ can be retrieved by taking the limit  $\epsilon \rightarrow 0$.

  The power of convolution as a regularization technique stems from the theorem:
\begin{quote}{\bf Theorem 2} \label{theo:2}
{\it The convolution $(D*\rho)(x)$ of a distribution $D \in \mathcal{D}'$ by a function $\rho \in \mathcal{D}$ is a $\mathcal{C}^\infty$ function in the variable $x$ \emph{\cite[p.\,465]{CHOQU1982-}}.}
\end{quote}
Regularized functions $f * \rho_\epsilon$ and distributions $D \ast \rho_\epsilon$ can therefore be freely multiplied and differentiated.  Moreover, the mollified sequence $D_\epsilon = D \ast \rho_\epsilon$ provides a representative sequence of the type \eqref{sch:2} of any distribution $D \in \mathcal{D}'$.

\section{Colombeau generalized functions} 
\label{col:0} \setcounter{equation}{0}

A Colombeau algebra $\mathcal{G}$ is an associative differential algebra in which multiplication, differentiation, and integration are similar to those of $\mathcal{C}^\infty$ functions.  Colombeau and others have introduced a number of variants of $\mathcal{G}$ but all `Colom\-beau algebras' have in common one essential feature: The $\mathcal{C}^\infty$ functions are a faithful differential subalgebra of $\mathcal{G}$, a feature that Colombeau discovered  to be essential to overcome Schwartz's multiplication-impossibility theorem. 

With hindsight it is easy to understand why:  If we suppose that $\mathcal{G}$ is an algebra  containing the distributions and such that all its elements can  be freely multiplied and differentiated just like $\mathcal{C}^\infty$ functions (i.e., in a way respecting commutativity, associativity, and the Leibniz rule), then $\mathcal{C}^\infty$ must be a subalgebra of $\mathcal{G}$ because $\mathcal{C}^\infty \subset \mathcal{D}'$. Thus, to define $\mathcal{G}$, it suffices to start from a differential algebra $\mathcal{E}$ containing the distributions, and then to define $\mathcal{G}$ as a subalgebra of $\mathcal{E}$ such that the embedding of the $\mathcal{C}^\infty$ functions in $\mathcal{G}$ is an identity.  In formulas: if $[g] \in \mathcal{G}$ represents an object $g$ embedded in $\mathcal{G}$, then for all $f \in \mathcal{C}^\infty$ we want that $[f]=f$, whereas for any other function or distribution $D \in \mathcal{D}'$ we may have $[D] \neq D$.

This simple observation gives a powerful hint for an elementary construction of $\mathcal{G}$ because by Definition \ref{defi:1} there is a one to one correspondence between any arbitrary distribution $D(x)$ and a class of weakly convergent sequence of $\mathcal{C}^\infty$ functions $D_\epsilon(x)$, and by Theorem \ref{theo:2} any representative of that class can be written as a convolution of the form \eqref{mul:2}.  Thus, the starting point is to consider for $\mathcal{E}$ the set of mollified sequencies\footnote{In the literature the notation  $(f_\epsilon)_\epsilon$ is often used here instead of $f_\epsilon$.  It emphasizes that $f_\epsilon$ is an element of $\mathcal{E}$ rather than just a representative sequence or a regularization.} 
\begin{eqnarray}
\label{col:1}
    \mathcal{E} \DEF  \Bigl\{
        f_\epsilon :  (\eta, x)  \mapsto  f_\epsilon(\eta,x)
                      \Bigr\},
\end{eqnarray}
which are $\mathcal{C}^\infty$ functions in the variable $x$ for any given \emph{Colombeau mollifier} $\eta$, and depend on the parameter $\epsilon \in ]0,1[$ through the scaled mollifier\footnote{In this and the next sections we take $\Omega=\mathbb{R}$ so that all integrals are $\int_{-\infty}^{+\infty}$.  The generalization to  $\Omega=\mathbb{R}^3$ is immediate, e.g., the scaled mollifier is $\eta(\vec{x}/\epsilon)/\epsilon^{3}$.}
\begin{eqnarray} \label{col:2}
     \eta_\epsilon(x) 
     \DEF \frac{1}{\epsilon} \eta\Bigl(\frac{x}{\epsilon}\Bigr),
     \qquad \text{normalized as} \qquad \int dy~\eta(y) = 1.
\end{eqnarray}
 The distributions $f \in \mathcal{D}'$ are then embedded in $\mathcal{E}$ as the convolution\footnote{This definition due to Colombeau differs by a sign from the usual definition \eqref{mul:2} of regularization.} 
\begin{eqnarray}
\nonumber
          f_\epsilon(x)  \DEF \eta_\epsilon(-x) \ast f(x)
                     &= \int dy~ \frac{1}{\epsilon}
                        \eta\Bigl(\frac{y-x}{\epsilon}\Bigr) ~f(y)\\
\label{col:3}   
                \,\, &= \int dz~\eta(z) ~f(x + \epsilon z),
\end{eqnarray}
where, in order to define $\mathcal{G} \subset \mathcal{E}$, the Colombeau mollifier $\eta$ may need to have specific properties in addition to the normalization \eqref{col:2}.

   To find these additional properties we have to study the embeddings of $\mathcal{C}^\infty$ functions.  We therefore calculate \eqref{col:3} for $f \in \mathcal{C}^\infty$, which enables to apply Taylor's theorem with remainder to obtain at once
\begin{eqnarray}
 \label{col:4}
    f_\epsilon(x) 
  &= f(x) \int dz~\eta(z) + ...\\ 
 \label{col:5}
  &+ \frac{\epsilon^n}{n!} f^{(n)}(x) \int dz~z^n \eta(z)
   + ...\\
 \label{col:6}
  &+ \frac{\epsilon^{(q+1)}}{(q+1)!}  \int dz~z^{q+1}\eta(z)
                           ~f^{(q+1)}(x+ \vartheta\epsilon z),
\end{eqnarray}
where $f^{(n)}(x)$ is the $n$-th derivative of $f(x)$, and $\vartheta \in ]0,1[$.  Since $f \in \mathcal{C}^\infty$ and $\eta$ has compact support, the integral in \eqref{col:6} is bounded so that the remainder is of order $\OOO(\epsilon^{q+1})$ at any fixed point $x$.

   Then, if following Colombeau the mollifier $\eta$ is chosen in the set
\begin{eqnarray} \label{col:7}
    \Bigl\{
    \int dz~\eta(z) = 1,
    \quad \text{and} \quad
    \int dz~z^n\eta(z) = 0,
    \quad \forall n=1,...,q \in \mathbb{N}
    \Bigr\},
\end{eqnarray}
all the terms in \eqref{col:5} with $n \in [1,q]$ are zero and we are left with
\begin{eqnarray}
 \label{col:8}
   \forall f \in \mathcal{C}^\infty, \qquad
    f_\epsilon(x) = f(x) + \OOO(\epsilon^{q+1}).
\end{eqnarray}
Therefore, provided $\eta$ is a Colombeau mollifier and $q$ can take any value in $\mathbb{N}$, it is possible to make the difference $f_\epsilon(x) - f(x)$ as small as we please even if $\epsilon \in ]0,1[$ is kept \emph{finite}.

   In terms of the embeddings \eqref{col:3} the condition $[f]=f$ insuring that the smooth functions are identically embedded in $\mathcal{G}$ is that $[f_\epsilon](x) = [f](x) = f(x)$ for all $f \in \mathcal{C}^\infty$.  Thus, comparing with \eqref{col:8}, we are led to consider the set $\mathcal{N}$ of the so-called \emph{negligible functions}, which correspond to the differences between the  $\mathcal{C}^\infty$ functions and their embeddings in $\mathcal{E}$, i.e., 
\begin{eqnarray} \label{col:9}
   \forall f \in \mathcal{C}^\infty,
   \quad \forall q \in \mathbb{N},
   \qquad f_\epsilon(x) -  f(x) = \OOO(\epsilon^q) \quad \in \mathcal{N}.
\end{eqnarray}
%
 
   To define $\mathcal{G}$ we need a prescription such that the differences \eqref{col:9} can be neglected, i.e., equated to zero in $\mathcal{G}$.  Moreover, for $\mathcal{G}$ to be an algebra, that prescription must be stable under multiplication.  That means that all elements $g \in \mathcal{G}$ have to have the property that any of their representatives $g_\epsilon \in \mathcal{E}$ multiplied by a negligible function are negligible.  In mathematical language,  $\mathcal{N}$ has to be an ideal of the subset $\{ g_\epsilon \} = \mathcal{E}_\Mt \subset  \mathcal{E}$ of all representatives of all elements of $\mathcal{G}$.  Or, in simple language, the negligible functions have to behave as the `function zero' when multiplying any function of $\mathcal{E}_\Mt$.   It is however very simple to characterize this subset:  Following Colombeau we call the elements of $\mathcal{E}_\Mt$ \emph{moderate (or multipliable) functions}, and we define
\begin{eqnarray} \label{col:10}
   \forall g_\epsilon \in \mathcal{E}_\Mt~:
   \qquad \exists N \in \mathbb{N}_0,
      \qquad \text{such that} \qquad 
   g_\epsilon(x) = \OOO(\epsilon^{-N}).
\end{eqnarray}
Indeed, as $q$ in \eqref{col:8} is as large as we please, and $N$ in \eqref{col:10} a fixed integer, the product of a negligible function by a moderate one will always be a negligible function: $\mathcal{N}$ is an ideal of $\mathcal{E}_\Mt$.  Moreover, the product of two moderate functions is still moderate:  They are \emph{multipliable}.

  For example, the Colombeau embeddings \eqref{col:3} of the $\delta$ and Heaviside functions are
\begin{eqnarray} \label{col:11}
     \delta_\epsilon(x) 
      = \frac{1}{\epsilon} \eta\Bigl(-\frac{x}{\epsilon}\Bigr),
    \qquad \text{and} \qquad
 \HU_\epsilon(x) = \int_{-\infty}^{x/\epsilon} dz~\eta(-z),
\end{eqnarray}
which are moderate functions with $N=1$ and $0$, respectively.  More generally, it can easily be proved using Schwartz's local structure theorem that:
\begin{quote}{\bf Theorem 3 (Colombeau local structure theorem)}\label{theo:3}
{\it Any distribution is locally a moderate (i.e., multipliable) generalized function \emph{\cite[p.\,61]{COLOM1984-}}, \emph{\cite[p.\,62]{GROSS2001-}}.}
\end{quote} 
Therefore,  $\mathcal{N} \subset (\mathcal{C}^\infty)_\epsilon \subset (\mathcal{C})_\epsilon \subset (\mathcal{D}')_\epsilon \subset \mathcal{E}_\Mt$.  It is also a matter of elementary calculation to verify that $\mathcal{E}_\Mt$ and $\mathcal{N}$ are algebras for the usual pointwise operations in $\mathcal{E}$.  Moreover, $\mathcal{E}_\Mt$ is a differential algebra, and it is not difficult to show that $\mathcal{E}_\Mt$ is the largest differential subalgebra (i.e., stable under partial differentiation) of $\mathcal{E}$ in which $\mathcal{N}$ is a differential ideal.

   The fact that $\mathcal{N}$ is an ideal of $\mathcal{E}_\Mt$ is the key to defining $\mathcal{G}$. Indeed, if we conventionally write $\mathcal{N}$ for any negligible function, then
\begin{eqnarray}
 \label{col:12}
   \forall g_\epsilon, h_\epsilon \in \mathcal{E}_\Mt, \qquad
   ( g_\epsilon + \mathcal{N} ) \cdot
   ( h_\epsilon + \mathcal{N} )
  = g_\epsilon \cdot h_\epsilon + \mathcal{N}.
\end{eqnarray}
Thus, it suffices to define the elements of $\mathcal{G}$ as the elements of $\mathcal{E}_\Mt$ modulo $\mathcal{N}$, i.e., to define the \emph{Colombeau algebra} as the quotient
\begin{eqnarray}
 \label{col:13}
        \mathcal{G} \DEF \frac{\mathcal{E}_\Mt}{\mathcal{N}}.
\end{eqnarray}
That is, an element $g \in \mathcal{G}$ is an equivalence class $[g] = [g_\epsilon + \mathcal{N}]$ of an element $g_\epsilon \in \mathcal{E}_\Mt$, which is called a \emph{representative} of the \emph{generalized function} $g$.  If `$\odot$' denotes multiplication in $\mathcal{G}$, the product $g \odot h$ is defined as the class of $g_\epsilon \cdot h_\epsilon$ where $g_\epsilon$ and $h_\epsilon$ are (arbitrary) representatives of $g$ and $h$; similarly $\rmD g$ is the class of $\rmD g_\epsilon$ if $\rmD$ is any partial differentiation operator.  Therefore, when working in $\mathcal{G}$, all algebraic and differential operations (as well as composition of functions, etc.) are performed component-wise at the level of the representatives $g_\epsilon$.

    $\mathcal{G}$ is an associative and commutative differential algebra because both $\mathcal{E}_\Mt$ and $\mathcal{N}$ are such.  The two main ingredients which led to its definition are the primacy given to $\mathcal{C}^\infty$ functions, and the use of Colombeau mollifiers for the embeddings. In fact, Colombeau proved that the set \eqref{col:7} is not empty and provided a recursive algorithm for constructing the corresponding mollifiers for all $q \in \mathbb{N}$.  He also showed \cite[p.\,169]{COLOM1992-} that the Fourier transformation provides a simple characterization of the mollifiers.  But, in the present paper as in most applications of the Colombeau algebras, the explicit knowledge of the form of the Colombeau mollifiers is not necessary:  It is sufficient to know their defining properties \eqref{col:7}.

    For example, let us verify that $\delta_\epsilon(x)$ given by \eqref{col:11} has indeed the sifting property expected for Dirac's $\delta$-function. Starting from \eqref{sch:1} and employing Taylor's theorem we can write
\begin{eqnarray}
\nonumber
    \BRA \delta_\epsilon | T \KET
 &= \int \delta_\epsilon(x) T(x)~dx 
  = \int_{-\infty}^\infty dx~ 
    \frac{1}{\epsilon} \eta\Bigl(-\frac{x}{\epsilon}\Bigr) T(x)
  = \int_{-\infty}^\infty dz~ \eta(-z) T(\epsilon z)\\
 \label{col:14}
 &= \int_{-\infty}^\infty dz~ \eta(-z)
  \Bigl( T(0) + \epsilon zT'(0) + \frac{(\epsilon z)^2}{2!}T''(0) + ... \Bigr),
\end{eqnarray}
Then, in Schwartz theory, we take the limit \eqref{sch:3}, i.e.,
\begin{eqnarray}
\label{col:15}
   \lim_{\epsilon \rightarrow 0} ~ \BRA \delta_\epsilon | T \KET  
  = T(0) + \OOO(\epsilon),
\end{eqnarray}
which is the expected result thanks to the normalization \eqref{col:2}.  However, in Colombeau theory, there is no need to take a limit to get the sifting property because in the development \eqref{col:14} the conditions \eqref{col:7} imply that all terms in $z^n$  with $1 < n < q+1$ are identically zero.  Thus
\begin{eqnarray}
\label{col:16}
\BRA \delta_\epsilon | T \KET  
  = T(0) + \OOO(\epsilon^{q+1}), \qquad \forall q \in \mathbb{N},
\end{eqnarray}
where the remainder is an element of $\mathcal{N}$ so that in $\mathcal{G}$ the sifting property of $\delta_\epsilon$ is an equality rather than a limit.  It is this kind of qualitative difference between the Schwartz and Colombeau theories which makes it possible in $\mathcal{G}$ to go beyond distribution theory.

\section{Interpretation and multiplication of distributions} 
\label{imd:0} \setcounter{equation}{0}

   To construct the Colombeau algebra we have been led to embed the distributions as the representative sequences $\gamma_\epsilon \in \mathcal{E}$ defined by \eqref{col:3} where $\eta$ is a Colombeau mollifier \eqref{col:7}.  We can therefore recover any distribution $\gamma$ by means of \eqref{sch:3}, i.e., as the equivalence class
\begin{eqnarray}
\label{imd:1}
        \gamma(T) \DEF \lim_{\epsilon \rightarrow 0} 
                 \int dx~\gamma_\epsilon(x) ~T(x),
                \qquad \forall T(x) \in \mathcal{D},
\end{eqnarray}
where $\gamma_\epsilon$ can be any representative of the class $[\gamma] = [\gamma_\epsilon + \mathcal{N}]$ because negligible elements are zero in the limit $\epsilon \rightarrow 0$.

   Of course, as we work in $\mathcal{G}$ and its elements get algebraically combined with other elements, there can be generalized functions $[g_\epsilon]$ different from the class $[\gamma_\epsilon]$ of an embedded distribution which nevertheless correspond to the {same} distribution $\gamma$.  This leads to the concept of \emph{association}, which is defined as follows,\footnote{In the literature the symbol $\approx$ is generally used for association.  We prefer to use $\asymp$ because association is not some kind of an `approximate' relationship, but rather the precise statement that a generalized function corresponds to a distribution.}
\begin{quote}{\bf Definition 2}
\label{defi:2}
{\it Two generalized functions $g$ and $h \in \mathcal{G}$, of respective representatives  $g_\epsilon$ and $h_\epsilon$, are said to be associated, and one write $g \ASS h$, iff }
\begin{eqnarray}
\label{imd:2}
                 \lim_{\epsilon \rightarrow 0} 
                 \int dx~\Bigl(g_\epsilon(x)-h_\epsilon(x)\Bigr) T(x) = 0,
                 \qquad \forall T(x) \in \mathcal{D}.
\end{eqnarray}
\end{quote}
Thus, if $g$ is a generalized function and $\gamma$ a distribution, the relation $g \ASS \gamma$ implies that $g$ admits $\gamma$ as `associated distribution,' and $\gamma$ is called the `distributional shadow' (or `distributional projection') of $g$.
 
   Objects (functions, numbers, etc.) which are equivalent to zero in $\mathcal{G}$, i.e., equal to $\OOO(\epsilon^q), \forall q \in \mathbb{N}$, are called `zero.'  On the other hand, objects associated to zero in $\mathcal{G}$, that is which tend to zero as $\epsilon \rightarrow 0$, are called `infinitesimals.'  Definition \eqref{imd:2} therefore means that two different generalized functions associated to the same distribution differ by an infinitesimal.

   The space of distributions is not a subalgebra of $\mathcal{G}$.  Thus we do not normally expect that the product of two distributions in $\mathcal{G}$ will be associated to a third distribution:  In general their product will be a genuine generalized function.

   For example, the square of Dirac's $\delta$-function, Eq.~\eqref{col:11}, which corresponds to $(\delta^2)_\epsilon(x) = (\delta_\epsilon)^2(x) = \epsilon^{-2}\eta^2(-x/\epsilon)$, has no associated distribution.  Indeed, making a Taylor development as in \eqref{col:14},
\begin{eqnarray}
\nonumber
    \lim_{\epsilon \rightarrow 0} 
    \BRA \delta_\epsilon^2 | T \KET 
 &= \lim_{\epsilon \rightarrow 0} \int_{-\infty}^\infty dx~ 
    \frac{1}{\epsilon^2} \eta^2\Bigl(-\frac{x}{\epsilon}\Bigr) T(x)
  = \lim_{\epsilon \rightarrow 0} \frac{1}{\epsilon} \int_{-\infty}^\infty dz~
    \eta^2(-z) T(\epsilon z)\\
\nonumber
 &= \lim_{\epsilon \rightarrow 0} \frac{1}{\epsilon}
    \int_{-\infty}^\infty dz~ \eta^2(-z)
    \Bigl( T(0) +\epsilon zT'(0)  +\frac{(\epsilon z)^2}{2!}T''(0) + ...\Bigr)\\
\label{imd:3}
 &= \lim_{\epsilon \rightarrow 0} \frac{T(0)}{\epsilon}
    \int_{-\infty}^\infty dz~ \eta^2(-z)
    + T'(0)\int_{-\infty}^\infty dz~ z\eta^2(-z)
  = \infty.
\end{eqnarray}
But, referring to \eqref{col:10}, $(\delta^2)_\epsilon(x)$ is a moderate function with $N=2$.  The square of $\delta(x)$ makes therefore sense in $\mathcal{G}$ as a `generalized function' with representative $(\delta^2)_\epsilon(x)=\eta^2(-x/\epsilon)/\epsilon^2$.  Moreover, its point-value at zero, $\eta^2(0)/\epsilon^2$, can be considered as a `generalized number.'  

   On the other hand, we have in $\mathcal{G}$ elements like the $n$-th power of Heaviside's function, Eq.~\eqref{col:11}, which has an associated distribution but is such that $[\HU^n](x)\neq[\HU](x)$ in $\mathcal{G}$, whereas $\HU^n(x) = \HU(x)$ as a distribution in $\mathcal{D}'$.  Similarly, we have $[x]\odot[\delta](x)\neq 0$ in $\mathcal{G}$, whereas $x\delta(x)=0$ in $\mathcal{D}'$.  In both cases everything is consistent:  Using \eqref{imd:2} one easily verifies that indeed $[\HU^n](x) \ASS [\HU](x)$ and $[x]\odot[\delta](x)\ASS 0$. 

  These differences between products in $\mathcal{G}$ and in $\mathcal{D}'$ stem from the fact that distributions embedded and multiplied in $\mathcal{G}$ carry along with them infinitesimal information on their `microscopic structure.'  That information is necessary in order that the products and their derivatives are well defined in $\mathcal{G}$, and is lost when the factors are identified with their distributional projection in $\mathcal{D}'$.  For example, since $[\HU^n](x)\neq[\HU](x)$ the inconsistencies displayed in \eqref{mul:1} do not arise in $\mathcal{G}$. Nevertheless, if at some point of a calculation it is desirable to look at the intermediate results from the point of view of distribution theory, one can always use the concept of association to retrieve their distributional content.  In fact, this is facilitated by a few simple formulas which easily derive from the definition \eqref{imd:2}.  For instance,
\begin{eqnarray}
\label{imd:4}
   \forall f_1,\forall f_2 \in \mathcal{C}
   \qquad &\Rightarrow \qquad
   [f_1]\odot[f_2] \ASS [f_1\cdot f_2],\\
%
\label{imd:5}
   \forall f \in \mathcal{C}^\infty,\forall \gamma \in \mathcal{D}'
   \qquad &\Rightarrow \qquad 
   [f]\odot[\gamma] \ASS [f\cdot \gamma],\\
%
\label{imd:6}
   \forall \gamma_1,\forall \gamma_2 \in \mathcal{D}'
   \qquad &\Rightarrow \qquad 
   [\gamma_1]\odot[\gamma_2] \not\ASS [\gamma_1\cdot \gamma_2],\\
%
\label{imd:7}
      \forall g_1,\forall g_2 \in \mathcal{G}, \qquad
      g_1 \ASS g_2 \qquad &\Rightarrow
                     \qquad \rmD^\alpha g_1 \ASS \rmD^\alpha g_2.
\end{eqnarray}
For example, applying the last equation to $[\HU^2](x) \ASS [\HU](x)$ one proves the often used distributional identity $2[\delta](x)[\HU](x) \ASS [\delta](x)$.

   In summary, one calculates in $\mathcal{G}$ as in $\mathcal{C}^\infty$ by operating on the representatives $g_\epsilon \in \mathcal{E}$ with the usual operations $\{ +,-,\times,d/dx\}$.  The distributional aspects, if required, can be retrieved at all stages by means of association.  However, as will be seen in the applications, it is possible in many cases to set aside the concept of distributions and to replace it by the more general and flexible one of $\mathcal{G}$-functions.

\section{Applications to hydrodynamics} 
\label{hyd:0} \setcounter{equation}{0}

   Distributions and their applications are common place in most areas of physics and engineering.  It is well known how to evaluate a $\delta$-function, how to calculate with piecewise continuous functions, etc.  In such applications there is no essential difference between Schwartz distributions and Colombeau functions.  This is because most characteristics distinguishing $\delta$-sequencies like for example those of Eq.~\eqref{sch:6} have no effect in such applications since the only things that matter when they are evaluated on test functions is their sifting property and their integral, which is normalized to one.  These characteristics are however relevant when distributions are multiplied and evaluated at the same $x$ in expressions such as $\delta^2(x)$, $\HU(x)\delta(x)$, $\HU_1(x)\HU_2(x)$, $\delta_1(x)\delta_2(x)$, $\HU_1(x)\delta_2(x)$, etc., where the indices 1 and 2 refer to different $\delta$-functions such that $\HU_i'=\delta_i$.

  For instance, products of the type $\HU_1(x)\HU_2(x)$ and $\HU_1(x)\delta_2(x)$ arise in the study of the propagation and interaction of shock waves, such as those occurring in strong collisions between projectiles and armor, a subject that has been extensively studies by Colombeau and his collaborators (Ref.~\cite{COLOM1992-} and references therein).  Shock waves induce sudden and large variations of physical quantities, for example the density, on a distance comparable to only a few times the average distance between molecules.  An Heaviside step function would therefore appear to be an excellent approximation of that behavior, which is indeed often the case.  However, in strong shocks during which a phase transition (e.g., from elastic to plastic) occurs this approximation is insufficient.  For instance, a typical combination of distributions arising in such a case are products like
\begin{eqnarray} \label{hyd:1}
                     \HU_1(x)\HU_2'(x) \ASS \alpha \delta(x),
\end{eqnarray}
where $\delta$ is $\HU_1'$ or $\HU_2'$.  It then turns out that whereas $\alpha=1/2$ in the simple case $\HU_1=\HU_2$, measurements show that $\alpha$ can be anything between $0.05$ and $0.95$, see \cite[p.\,43--48]{COLOM1992-}, implying that the `microscopic profiles' of $\HU_1$ and $\HU_2$ are very different at the jump, i.e., $\delta_1 \neq \delta_2$.

  A lesson from this application is that by assuming that physically relevant distributions such as $\HU$ and $\delta$ are elements of $\mathcal{G}$ one gets a picture that is much closer to reality than if they are restricted to $\mathcal{D}'$.  In fact, this lesson applies not just to numerical modeling and applied physics but also to mathematics and theoretical physics.

  Consider for example one of the simplest nonlinear partial differential equations, the inviscid Burger's equation of hydrodynamics, which in $\mathcal{G}$ can be written in two ways
\begin{eqnarray} \label{hyd:2}
         u_t + u u_x = 0, ~~~ ~~~ \text{or} ~~~ ~~~  u_t + u u_x \ASS 0.
\end{eqnarray}
Both equations have a traveling wave solutions of the type $u(x,t) = (u_2-u_1)  \HU(x-ct) + u_1$, which using the identity $2\HU \HU'=\HU'$ yields a jump velocity of $c=(u_1+u_2)/2$.  But, if multiplied by $u$, the first equation has an additional solution which turns out to be inconsistent with the first one, so that it has in fact no solutions.  This is however not the case with the second equation because whereas multiplication is compatible with equality in $\mathcal{G}$, it is not compatible with association.  Therefore, the distinction between $=$ and $\ASS$ automatically insures that the physically correct solution is selected, a distinction that can be made in analytical as well as in numerical calculations by using a suitable algorithm \cite{COLOM1992-}.

\section{Point charges in classical electrodynamics} 
\label{poi:0} \setcounter{equation}{0}

   It is well known that classical electrodynamics is basically a continuum theory.  Nevertheless, point charges are very often considered --- even in the most elementary introductory lectures.  Whenever a contradiction arises the difficulty is then set aside:  Infinite quantities are discarded or renormalized, self-interactions terms in Lagrangians are ignored, etc.


    It is however possible to consistently introduce point charges through distribution theory, and to deal with them successfully, at least as long as the electromagnetic fields, currents, and charge densities interpreted as distributions are not multiplied.\footnote{It is also possible to deal in specific cases with problems that are non-linear in the fields.  But this requires ad hoc prescriptions, such as defining the energy-momentum tensor, which is quadratic in the fields, directly as a distribution.  See, e.g., \cite{TAYLO1956-,ROWE-1978-}.}  Since distributions are a subspace of $\mathcal{G}$ nothing fundamentally new has to be invented, and the electromagnetic field distributions defined in the context of Schwartz distribution theory can just as well be used in the $\mathcal{G}$-setting.  This provides a `standard methodology' in which these distributions are simply embedded in $\mathcal{G}$ by means of a Colombeau regularization.  But, since the Colombeau regularization has specific properties that standard regularizations do not have, it is possible to develop a more powerful and convenient methodology which takes these properties explicitly into account.  This methodology, based on a suitable generalized function $\UPS$, has already been presented in References~\cite{GSPON2004D} and \cite{GSPON2007C}, albeit with only a simplified and intuitive justification of its validity.  In the following we are going to give a rigorous mathematical foundation to it, starting with a presentation of the standard methodology, and then moving in successive steps to the improved one.

   The basic idea in distribution theory is to replace the classical Coulomb potential $e/r$ of a point charge by the weak limit of the sequence of distributions \cite[p.\,144]{TEMPL1953-}, \cite[p.\,51]{SCHUC1991-},
\begin{eqnarray} \label{poi:1}
  \phi(r) \DEF \frac{e}{r} \HU_{\a}(r),
\qquad \text{where} \qquad
             \HU_{\a}(r) \DEF \lim_{a \rightarrow 0} \HU(r-a),
\end{eqnarray}
where $e$ is the electric charge of an electron at rest at the origin of a polar coordinate system, and $r = |\vec{r}\,|$ the modulus of the radius vector.\footnote{We write $\a$ rather than $a$ in subscript to emphasize that $a$ is not an index but the parameter of a limiting sequence.}  Consistent with Schwartz's local structure theorem, $\phi(r)$ is the derivative of $e\lim_{a \rightarrow 0} \log(r/a) \HU(r-a)$, a $\mathcal{C}^0$ function $\forall r \geq 0$.  The cut-off $a > 0$ insures that $\phi(r)$ is a well defined piecewise continuous function for all $r \geq 0$, whereas the classical Coulomb potential $e/r$ is defined only for $r > 0$.  It is then readily verified, using \eqref{sch:3}, that \eqref{poi:1} is a distribution.  Indeed, since $\Omega = \mathbb{R}^3$,
\begin{eqnarray} \label{poi:2}
   \forall T \in \mathcal{D}, \qquad 
   \iiint_{\mathbb{R}^3} d^3r~ \phi(r) T(r) =
  4\pi e \int_0^\infty dr~ r T(r) \quad \in \mathbb{R},
\end{eqnarray}
because $T \in \mathcal{D}$ has compact support so that the integral is bounded.

   The next step, in order to be able to calculate the field $\vec{E} = -\vec{\nabla} \phi$ and the charge density $4\pi\rho = \vec{\nabla} \cdot \vec{E}$, is to represent $\phi$ by a mollified sequence.  Taking for it the Colombeau form \eqref{col:3} the Coulomb potential is at once embedded in $\mathcal{G}$ as the representative sequence
\begin{eqnarray} \label{poi:3}
  \phi_\epsilon(r) = \bigl(\frac{e}{r} \HU_{\a}\bigr)_\epsilon(r)
             = e \lim_{a \rightarrow 0}
                 \int_{\frac{a-r}{\epsilon}}^\infty dz~
                 \frac{\eta(z)}{r+\epsilon z},
                 \qquad \forall r \geq 0.
\end{eqnarray}
One then trivialy verifies that the distributional Coulomb potential \eqref{poi:1} can be recovered by letting $\epsilon \rightarrow 0$ as in \eqref{imd:2}, i.e., 
\begin{eqnarray} \label{poi:4}
  \phi_\epsilon(r) \ASS \frac{e}{r} \HU_{\a}(r) \DEF \phi(r),
\end{eqnarray}
which reverts to the classical Coulomb potential $e/r$ as $a \rightarrow 0$.

  Calculating the embedded Coulomb field is now straightforward because the embedded potential $(e\HU_{\a}/r)_\epsilon$ is $\mathcal{C}^\infty$ in the variable $r$.  It becomes
\begin{eqnarray} \label{poi:5}
  \vec{E}_\epsilon(\vec{r}\,) = -\vec{\nabla} \phi_\epsilon(r)
              = e \lim_{a \rightarrow 0} \Bigl(
            \int_{\frac{a-r}{\epsilon}}^\infty dz~
                            \frac{\eta(z)}{(r+\epsilon z)^2}
          - \frac{1}{\epsilon a}\eta\bigl(\frac{a -r}{\epsilon}\bigr)
                        \Bigr) \vec{u},
\end{eqnarray}
where $\vec{u} = \vec{\nabla} r$ is the unit vector in the direction of $\vec{r}$.  Introducing the notation
\begin{eqnarray} \label{poi:6}
   \delta_{\a}(r) \DEF \lim_{a \rightarrow 0} \delta(r-a),
   \qquad \text{so that} \qquad
   \lim_{a \rightarrow 0}
   \frac{1}{\epsilon a}\eta\bigl(\frac{a -r}{\epsilon}\bigr)
 = \bigl(\frac{1}{a} \DUP_{\a}\bigr)_\epsilon(r),
\end{eqnarray}
this electric field can be written in the more convenient form
\begin{eqnarray} \label{poi:7}
  \vec{E}_\epsilon(\vec{r}\,) = e \Bigl(
            \bigl(\frac{1}{r^2} \HU_{\a}\bigr)_\epsilon(r)
          - \bigl(\frac{1}{a} \DUP_{\a}\bigr)_\epsilon(r)
                        \Bigr) \vec{u}.
\end{eqnarray}
By an appeal to test functions we easily verify that the field $\vec{E}_\epsilon$ is a distribution, and that the $\DUP$-function in \eqref{poi:7} gives a nul contribution when evaluated on a test function.  Thus
\begin{eqnarray} \label{poi:8}
  \vec{E}_\epsilon(\vec{r}\,) \ASS \frac{e}{r^2} \HU_{\a}(r) \vec{u}
                          \DEF \vec{E}(\vec{r}\,),
\end{eqnarray}
where $\vec{E}(\vec{r}\,)$ is the distributional Coulomb field which in the limit $a \rightarrow 0$ yields the classical Coulomb field $e\vec{r}/r^3$.  Therefore, the distribution $\vec{E}(\vec{r}\,)$ associated to the $\mathcal{G}$-function $\vec{E}_\epsilon$, i.e., its `shadow' obtained by projecting it on $\mathcal{D}'$, does not contain the $\DUP$-function contribution on the right of \eqref{poi:7}.

  To get the Coulomb charge density we have to calculate the divergence of \eqref{poi:5}.  In standard distribution theory one would then ignore the term on the right because it corresponds to a $\DUP$-function which, as we have just seen, gives no contribution when evaluated on a test function.  However, in $\mathcal{G}$, this term cannot be ignored if we subsequently calculate quantities in which $\vec{E}_\epsilon$ is a factor in a product.  Calculating $\rho_\epsilon$ is therefore somewhat laborious, but still elementary. It yields, using $\vec{\nabla} \cdot \vec{u} = 2/r$,
\begin{eqnarray}
\nonumber
  4\pi\rho_\epsilon(r) &= \vec{\nabla} \cdot \vec{E}_\epsilon(\vec{r}\,)
              = e \lim_{a \rightarrow 0} \Bigl( ~
        \frac{2}{r} \int_{\frac{a-r}{\epsilon}}^\infty dz~
                            \frac{\eta(z)}{(r+\epsilon z)^2}
        - \frac{2}{\epsilon a r}\eta\bigl(\frac{a -r}{\epsilon}\bigr)\\
 \label{poi:9}
        &+ \frac{1}{\epsilon a^2}\eta\bigl(\frac{a -r}{\epsilon}\bigr)
        -2 \int_{\frac{a-r}{\epsilon}}^\infty dz~
                            \frac{\eta(z)}{(r+\epsilon z)^3}
        + \frac{1}{\epsilon^2 a}\eta'\bigl(\frac{a -r}{\epsilon}\bigr)
                       ~ \Bigr).
\end{eqnarray}
This expression can be rewritten in the less cumbersome form
\begin{eqnarray}
\nonumber
  4\pi\rho_\epsilon(r) &= e \Bigl( ~
        \frac{2}{r} \bigl(\frac{1}{r^2} \HU_{\a}\bigr)_\epsilon(r)
                 -2 \bigl(\frac{1}{r^3} \HU_{\a}\bigr)_\epsilon(r)\\
 \label{poi:10}
        &+ \bigl(\frac{1}{a^2} \DUP_{\a}\bigr)_\epsilon(r)
         - \frac{2}{r}\bigl(\frac{1}{a} \DUP_{\a}\bigr)_\epsilon(r)
         - \bigl(\frac{1}{a}  \DUP'_{\a}\bigr)_\epsilon(r)
                       ~ \Bigr),
\end{eqnarray}
where we have put the two Heaviside-function terms on the first line and the three Dirac-function ones on the second.  This charge density is of course much more complicated than the single three-dimensional $\delta$-function which is associated to it in the standard distributional formalism.  But it is the correct result, and a typical example of how quickly calculations become complicated when the infinitesimal details of distributions are fully taken into account.  To calculate the distributional shadow associated to \eqref{poi:10} we remark that the two Heaviside-function terms on the first line cancel each other when $\epsilon \rightarrow 0$, and that, in this limit, the representatives of the $\DUP$-functions on the second line become genuine $\DUP$-functions.  Thus, evaluated on a test function $T$, the first two $\delta$-terms give
\begin{eqnarray} \label{poi:11}
  \int_0^\infty dr ~ r^2 \bigl( \frac{1}{a^2} - \frac{2}{r}\frac{1}{a} \bigr)
         \DUP(r-a) T(r)  = - T(a),
\end{eqnarray}
whereas the $\delta'$-term gives, using integration by parts,
\begin{eqnarray} 
\nonumber
  \int_0^\infty dr ~ r^2 \bigl(- \frac{1}{a}\bigr) \DUP'(r-a) T(r)  
     &= - \int_0^\infty dr ~  \bigl(-\frac{r^2}{a} T(r)\bigr)' \DUP(r-a)\\
\label{poi:12}
     &= 2T(a) + a T'(a).
\end{eqnarray}
Thus, adding \eqref{poi:11} and \eqref{poi:12}, and passing to the limit $a \rightarrow 0$, we get the test-function-evaluation $T(0)$ so that
\begin{eqnarray} \label{poi:13}
  \rho_\epsilon(r) \ASS \frac{e}{4\pi r^2} \DUP_{\a}(r) \DEF \rho(r),
\end{eqnarray}
which yields the classical point-charge density $ e\delta^3(r) = e \delta(r)/4\pi r^2$ as $a \rightarrow 0$.

\section{Point charges in  $\UPS$-methodology}
\label{ups:0} \setcounter{equation}{0}

As we have seen in the previous section, calculations with distributions embedded in $\mathcal{G}$ can easily become tedious.  However, this is mainly because the representatives used for $\phi$, $\vec{E}$, and $\rho$ are mollified embeddings of sequencies of distributions, such as \eqref{poi:3} for $\phi$, in which the full properties \eqref{col:7} of the Colombeau mollifiers have not yet been taken into account.  Indeed, instead of \eqref{poi:3}, the Coulomb potential can actually be written in terms of a $\mathcal{G}$-function $\UPS$ so that it reads \cite{GSPON2004D}
\begin{eqnarray} \label{ups:1}
 \phi(r) =  \frac{e}{r} \UPS(r),
     \qquad \text{where}\qquad
     \frac{1}{r}  \UPS(r) =
          {
          \cases{
                      0          &for\quad   $r = 0$,\\
                      r^{-1}     &for\quad   $r > 0$,
                }
          }
\end{eqnarray}
which has the advantage that most calculations can be made as if $1/r$ and $\UPS(r)$ were ordinary functions.

  In fact, a representative of the $\mathcal{G}$-function $\UPS$ is provided by the embedding of the $\HU_{\a}$ distribution, i.e.,
\begin{eqnarray} 
\label{ups:2}
   \UPS(r) &\DEF \lim_{a \rightarrow 0} \lim_{\epsilon \rightarrow 0} 
   \bigl(\HU_{a}\bigr)_\epsilon(r) 
 = \lim_{a \rightarrow 0} \lim_{\epsilon \rightarrow 0} 
   \int_{\frac{a-r}{\epsilon}}^\infty dz~ \eta(z),
\end{eqnarray}
and a representative of its derivative $\UPS'$ by the embedding of the $\delta_{\a}$ distribution,
\begin{eqnarray} 
\label{ups:3}
   \UPS'(r) &\,= \lim_{a \rightarrow 0} \lim_{\epsilon \rightarrow 0} 
   \bigl(\DUP_{a}\bigr)_\epsilon(r)
 = \lim_{a \rightarrow 0} \lim_{\epsilon \rightarrow 0} 
   \frac{1}{\epsilon} \eta\Bigl(\frac{a-r}{\epsilon}\Bigr).
\end{eqnarray}
This implies that $\UPS$ has properties similar to Heaviside's step-function, and $\UPS'$ to those of Dirac's $\delta$-function, with the fundamental difference that they are $\mathcal{G}$-functions which can be freely multiplied, differentiated, integrated, and combined with any $\mathcal{C}^\infty$ function.  Moreover, the combination $r^{-n}\UPS(r)$ can be given a precise meaning, and the number $0$ in $\mathcal{G}$, i.e., $\OOO (\epsilon^q), \forall q \in \mathbb{N}$, assigned to its point-value at $r=0$.  More precisely, the essential property of $\UPS$ is 
\begin{eqnarray} \label{ups:4}
    \bigl(\frac{1}{r^n}  \UPS \bigr)(r)
        = \frac{1}{r^n}  \UPS(r) =
          {
          \cases{
                      0          &for\quad   $r = 0$,\\
                      r^{-n}     &for\quad   $r > 0$,
                }
          }
\end{eqnarray}
in which the symbol $r^{-n}\UPS(r)$ is interpreted as a whole such that the point-value $r^{-n}\UPS(r)\bigr|_{r=0}=0$ is well defined.  This property has many practical consequencies: For example, it enables to write the potential $\phi(r)$ as in \eqref{ups:1}, and it allows to freely associate the powers of $r$ in products such as  
\begin{eqnarray} \label{ups:5}
 r^p \Bigl( \frac{1}{r^n}\UPS(r) \Bigr) \Bigl( \frac{1}{r^m}\UPS(r) \Bigr)
   = \frac{r^p}{r^{n+m}}\UPS^2(r).
\end{eqnarray}

    The detailed proof of \eqref{ups:4}, which is given in \cite{GSPON2008B}, is not very complicated but somewhat lengthy.  However, it is intuitively evident that the $\mathcal{G}$-identity \eqref{ups:4} is simply the equivalent modulo $\OOO(\epsilon^q), \forall q \in \mathbb{N},$ of the distributional identity
\begin{eqnarray} \label{ups:6}
    \bigl(\frac{1}{r^n}  \HU_{\a} \bigr)(r) 
        = \frac{1}{r^n}  \HU_{\a}(r) =
          {
          \cases{
                      0          &for\quad   $r = 0$,\\
                      r^{-n}     &for\quad   $r > 0$,
                }
          }
\end{eqnarray}
where the effect of $\HU_{\a}$ is to `cut-off' the singularity at $r=0$ as in \eqref{poi:1}.

   Equation \eqref{ups:4} enables to work in $\mathcal{G}$ with a much larger class of test functions than just the usual $T(\vec{x}\,) = T(r,\theta, \phi) \in \mathcal{D}(\mathbb{R}^3)$ of standard distribution theory.  This class of `generalized test functions' is
\begin{eqnarray} \label{ups:7} 
  \mathcal{F} \DEF \Bigl\{  F(\vec{x}\,) = \frac{T(r,\theta, \phi)}{r^n},
   \quad \forall T \in \mathcal{D}(\mathbb{R}^3),
   \quad \forall n \in \mathbb{N}_0   \Bigr\}.
\end{eqnarray}
Due to \eqref{ups:4}, the functions $F(r,\theta, \phi) \in \mathcal{F}(\mathbb{R}^3)$ are well defined at $r=0$ when multiplied by $\UPS$ or any of its derivatives.  This enables to integrate the $\mathcal{G}$-functions $\UPS/r^n$, $\UPS'/r^n$, etc., which arise when calculating the electromagnetic fields of point charges and their derivatives, as well as any algebraic combination of them \cite{GSPON2008B}.   We will however content ourselves with just a few basic integration formulas, which are proved in the Appendix, i.e.,
%
%
\begin{eqnarray}
 \label{ups:8}
  \int_0^\infty dr~ \UPS(r)      F(r) &= \lima \int_a^\infty dr~ F(r),\\
 \label{ups:9}
  \int_0^\infty dr~ \UPS^2(r)    F(r) &= \lima \int_a^\infty dr~ F(r),\\
 \label{ups:10} 
  \int_0^\infty dr~ \UPS'(r)     F(r) &= \lima F(a),\\
 \label{ups:11} 
  \int_0^\infty dr~ \UPS(r)\UPS'(r) F(r) &= \lima \frac{1}{2} F(a),\\
 \label{ups:12}
  \int_0^\infty dr (\UPS')^2(r)T(r) 
         &= \lim_{\epsilon \rightarrow 0} C_{[0]}\frac{T(0)}{\epsilon}
                                        + C_{[1]} {T'(0)},
\end{eqnarray}
where $F(r) \in \mathcal{F}$, and $T(r) \in \mathcal{D}$.  As expected, the $\mathcal{G}$-functions $\UPS$ and $\UPS'$ have properties similar to the distributions $\HU$ and $\DUP$.  Also, whereas formulas (\ref{ups:8}--\ref{ups:11}) do not depend of the particular representative of $\UPS$ (i.e., on the shape of $\eta$ provided $\int_{-1}^{+1} \eta(x)\,dx =1$), formula \eqref{ups:12} explicitly depends on it because  
\begin{eqnarray}
\label{ups:13} 
  C_{[0]} = \int_{-\infty}^{+\infty} dx~   \eta^2(-x),
            \qquad \text{and} \qquad
  C_{[1]} =  \int_{-\infty}^{+\infty} dx~ x \, \eta^2(-x),
\end{eqnarray}
where $C_{[1]} = 0$ if $\eta$ is even.  This kind of undeterminedness is an intrinsic feature of the nonlinear context:  Products of $\mathcal{G}$-functions lead to results which in general depend on the form of their representatives, a form which is determined by the physical problem.

   To conclude this section we recalculate the Coulomb field and charge density in the $\UPS$-formalism, that is starting from the potential \eqref{ups:1}, which being a $\mathcal{C}^\infty$ expression permits to calculate as in elementary vector analysis.  Thus, since $\vec{\nabla} r = \vec{u}$,
\begin{eqnarray} \label{ups:14}
  \vec{E}(\vec{r}\,)  = -\vec{\nabla} \phi(r)
                      = e \Bigl(
                \frac{1}{r^2} \UPS(r)
              - \frac{1}{r}   \UPS'(r)
                          \Bigr) \vec{u},
\end{eqnarray}
which is fully equivalent to \eqref{poi:7}.   The calculation of $\rho(r)$ is also elementary, and leads to a greatly simplified result.  Indeed, as $\vec{\nabla} \cdot \vec{u} = 2/r$,
\begin{eqnarray}
\nonumber
  4\pi\rho(r)  &= \vec{\nabla} \cdot \vec{E}(\vec{r}\,)
     = e \Bigl( \frac{1}{r^2}\UPS(r) - \frac{1}{r}\UPS'(r) \Bigr)\frac{2}{r}\\
\label{ups:15}
     &+ e \Bigl(-\frac{2}{r^3}\UPS(r)  + \frac{1}{r^2}\UPS'(r)
               + \frac{1}{r^2}\UPS'(r) - \frac{1}{r}  \UPS''(r) \Bigr)
      = - e \frac{1}{r} \UPS''(r),
\end{eqnarray}
which is much simpler than \eqref{poi:10}, and is easily seen to be associated to the usual three-dimensional charge-density because $\UPS''(r)/r \ASS -\delta(r)/r^2$ in $\mathbb{R}^3$.  Moreover, this expression has the virtue of clearly showing the `origin' of the charge density: The $\UPS(r)$ factor in the potential \eqref{ups:1}.

\section{Point-charge self-energy in $\mathcal{D}'$ and $\mathcal{G}$}
\label{sen:0} \setcounter{equation}{0}

Now that we have derived the Colombeau generalized functions corresponding to the Coulomb potential, field, and charge distributions, it is of interest to show that the `extra' $\delta$-like terms in \eqref{poi:7} and \eqref{ups:14} --- which disappear when considering these distributions in $\mathcal{D}'$ rather than in $\mathcal{G}$ --- are physically significant.  To do this we calculate the self-energy of a point charge using the electric field $\vec{E}$ defined according to three theories:  The classical theory, distribution theory, and the $\mathcal{G}$ theory, but using in all three cases  the same self-energy expression derived from the Maxwell energy-momentum tensor.  That is, in the classical theory in which $\vec{E}(\vec{r}\,)$ is just the Coulomb field $e\vec{r}/r^3$, the integral
\begin{eqnarray} \label{sen:1}
 U_{{\rm self}} \DEF \frac{1}{8\pi} \iiint_{\mathbb{R}^3} d^3r~ \vec{E}^2 
                    = \frac{1}{2} \int_0^\infty dr~ r^2\frac{e^2}{r^4}
                    = \frac{e^2}{2}\lim_{r \rightarrow 0} \frac{1}{r} = \infty.
\end{eqnarray}

    In distribution theory we take for the Coulomb field the distribution $\vec{E}(\vec{r}\,) = e\HU_{\a}\vec{r}/r^3$ defined by \eqref{poi:8}.  Then, apart from expressing the self-energy $U_{{\rm self}}(T)$ as a function of the cut-off $a$, we still have the same divergent result
\begin{eqnarray} \label{sen:2}
 U_{{\rm self}}(1) = \frac{1}{8\pi}  \BRA \vec{E}^2 | 1 \KET
                   = \frac{1}{8\pi} \iiint_{\mathbb{R}^3} d^3r~ \vec{E}^2
                   = \frac{e^2}{2}\lim_{a \rightarrow 0} \frac{1}{a} = \infty,
\end{eqnarray}
even if $\vec{E}^2$ is evaluated on a test-function $T\neq 1$.  Thus, whereas the distribution \eqref{poi:8} is meaningful for all $r \geq 0$, and gives sensible results for expressions linear in $\vec{E}$ evaluated on any test-function, it does not give a sensible result for the self-energy, which is quadratic in  $\vec{E}$.  In particular, it is not possible to take the limit $a \rightarrow 0$ which is mandatory for having a \emph{point} charge.

    We now calculate the self-energy in $\mathcal{G}$, where the square of $\vec{E}_\epsilon$ is well defined.  With the Coulomb field expressed as \eqref{ups:14},  the self-energy is 
\begin{eqnarray}
\nonumber
 U_{{\rm self}} &= \frac{1}{8\pi} \iiint_{\mathbb{R}^3} d^3r~ \vec{E}^2 
      = \frac{e^2}{2} \int_0^\infty dr~ r^2
           \Bigl( \frac{1}{r^2} \UPS(r) - \frac{1}{r} \UPS'(r) \Bigr)^2\\
 \label{sen:3}
      &= \frac{e^2}{2} \int_0^\infty dr~ 
\Bigl(\frac{1}{r^2}\UPS^2(r) - \frac{2}{r}\UPS(r)\UPS'(r) + (\UPS')^2(r) \Bigr),
\end{eqnarray}
where all terms could freely be multiplied, and the $r^2$ factor simplified with the $1/r^n$ factors, because everything is $\mathcal{C}^\infty$. Then,  after integration, the first two terms cancel each other exactly.  Indeed, due to the identity
\begin{eqnarray}
 \label{sen:4}
       \Bigl(-\frac{1}{r}\UPS^2(r)\Bigr)'
     = \frac{1}{r^2}\UPS^2(r) - \frac{2}{r}\UPS(r)\UPS'(r),
\end{eqnarray}
we get, integrating \eqref{sen:3} by parts,
\begin{eqnarray}
\label{sen:5}
 U_{{\rm self}}  = -\frac{1}{r}\UPS^2(r) \Bigr|_0^\infty
      + \frac{e^2}{2} \int_0^\infty dr~ (\UPS')^2(r),
\end{eqnarray}
where the first term is zero on account of \eqref{ups:4}, whereas the integral of the $(\UPS')^2$ term gives by means of \eqref{ups:12} the result
\begin{eqnarray}
\label{sen:6}
 U_{{\rm self}}  = \frac{e^2}{2} \lim_{\epsilon \rightarrow 0}
                    \frac{1}{\epsilon} \int_{-\infty}^{+\infty} dx~ \eta^2(-x).
\end{eqnarray}
Because of the cancellation of the first two terms in \eqref{sen:3}, only the square of the $\UPS'$ function, whose support is a \emph{point}, contributes to the self-energy.  This means that while the self-energy is infinite in the limit $\epsilon \rightarrow 0$, this energy is now `concentrated' at the location of the point charge rather than spread over the whole space surrounding it.

   Summarizing, when the self-energy is calculated in $\mathcal{G}$ rather than in $\mathcal{D}'$, the divergent classical Coulomb-field self-energy \eqref{sen:2} is canceled by the mixed term in the integral \eqref{sen:3}, and that cancellation is exact and independent of the limiting sequence $a \rightarrow 0$ which is implicit in the symbol $\UPS$.  The sole contribution to the self-energy comes then from the $\delta^2(r)$ term in that integral.  This yields the result \eqref{sen:6} which depends only on the shape of the mollifier $\eta$ and on the regularization parameter $\epsilon$, and which may be renormalized to a finite quantity such as the mass of the point charge.

We have therefore obtained the physically remarkable result that in the Colom\-beau algebra  --- in which the multiplication of distributions is a meaningful operation --- the self-energy of a point-charge is entirely located at the position of the charge, and solely due to the square of the $\UPS'(r)$ term in the electric field \eqref{ups:14}, which itself derives form the $\UPS(r)$ factor in the potential \eqref{ups:1}.

\section{Application of $\UPS$-methodology to linear problems}
\label{lin:0} \setcounter{equation}{0}

The final purpose of this paper is to confirm the methods used in papers \cite{GSPON2004D,GSPON2007C}, in which only linear problems were considered.  This is easily done by referring to the foregoing subsections, and by making the simplifications that are possible in that context.

Indeed, in such applications there are no products of distributions such as $\UPS^2$, $\UPS\UPS'$, or $(\UPS')^2$.  The only remaining integration formulas are \eqref{ups:8} and \eqref{ups:10}, which reduce to the usual equations defining the properties of the Heaviside and Dirac distributions provided one substitutes $\UPS(r) \rightarrow \HU(r)$ and $\UPS'(r) \rightarrow \delta(r)$.  However, as was noted in \cite{GSPON2004D}, the property $\UPS(r)|_{r=0}=\UPS(0)=0$ is necessary for the consistency of the formalism so that we keep the notation $\UPS$ for that generalized function.  On the other hand, there is no absolute necessity to distinguish between $\UPS'(r)$ and $\delta(r)$.  Thus, instead of $\UPS''(r)$, $\UPS'''(r)$, etc., one can systematically use the associated distributional expressions $-\delta(r)/r$, $2\delta(r)/r^2$, etc., that is, $(\UPS')^{(n)} \ASS (-1)^n~n!~\DUP(r)/r^n$.

Then, after all these simplifications, one may wonder why it is necessary to refer to generalized functions to justify the $\UPS$-formalism in the linear context.  The answer is that the simple rules introduced and used in \cite{GSPON2004D,GSPON2007C} --- which imply assigning the point value $0$ to $\UPS(0)$ and working with $\UPS$ and $\DUP$ as if they were $\mathcal{C}^\infty$ functions --- only make fully sense in a framework of generalized functions such as a Colombeau algebra.  For this reason, while it is possible to forget about most of the technicalities of that theory when using the $\UPS$-formalism in linear problems, it is important to appreciate that working with $\UPS'$ instead of $\delta$, as in Secs.~\ref{ups:0}--\ref{sen:0} of the present paper, is not much more complicated, and possibly less prone to mistakes.

Finally, and to conclude, it is perhaps important to stress that the Colombeau formalism highlights the non-unicity of the `microscopic' representations which at the end of a calculation give a physically meaningful result at the `macroscopic' level, i.e., when evaluated on a test-function.  This is why the methodologies that we labeled `standard' and `$\UPS$' are most probably not the final theories which at the `microscopic' level have a physically meaningful interpretation of their own:  This is possibly much more the case of the methodology first introduced by Frank R.~Tangherlini,\footnote{This method has been independently rediscovered by the author and by others.  See \cite{GSPON2004D} and references therein.} which has the virtue of associating the $\mathcal{G}$-function $\UPS$ to the discontinuity of a truly fundamental quantity, the absolute value $|\vec{r}\,|$ of the distance between a source-point and a test-point, rewritten as $|\vec{r}\,| = r \UPS(r)$ so that this discontinuity is properly taken into account \cite{TANGH1962-}.

\section{Acknowledgments}
\label{ack:0} \setcounter{equation}{0}

The author would like to thank Professor Hanno Ess\'en for his continuing encouragement and several suggestions which led to significant improvements of this paper.

\section{Appendix:  Proof of integration formulas}
\label{app:0} \setcounter{equation}{0}

To prove \eqref{ups:8} and \eqref{ups:9} it suffices to refer to \eqref{ups:4} which specifies that apart from the point $r=0$ the function $\UPS(r)$ can be identified with one.  Thus, for any $m$,
\begin{eqnarray} \label{app:1}
  \int_0^\infty dr~ \UPS^m(r) ~ F(r)
 = \lim_{a \rightarrow 0} \int_a^\infty dr~ F(r).
\end{eqnarray}

   To prove \eqref{ups:10} we integrate by parts its left-hand side, i.e., 
\begin{eqnarray} 
\label{app:2}
         \int_0^\infty dr~ \UPS'(r) F(r) 
      &= \UPS(r)F(r)\Bigr|_0^\infty
       -  \int_0^\infty dr~ \UPS(r)F'(r)\\
\label{app:3}
      &= - \lima \int_a^\infty dr~ F'(r)  = \lima F(a),
\end{eqnarray}
where \eqref{ups:4} and \eqref{app:1} were used in \eqref{app:2}, and $F(\infty)=0$ in \eqref{app:3}.
 
   Similarly, the identity $(\UPS^2)' = 2\UPS\UPS'$ enables to integrate \eqref{ups:11} by parts,
\begin{eqnarray} 
\label{app:4}
         \int_0^\infty dr~ \UPS(r) \UPS'(r) F(r) 
      &= \frac{1}{2}\UPS^2(r)F(r)\Bigr|_0^\infty
       - \frac{1}{2} \int_0^\infty dr~ \UPS^2(r)F'(r)\\
\label{app:5}
      &= - \lima \frac{1}{2} \int_a^\infty dr~ F'(r) 
       =   \lima \frac{1}{2}F(a),
\end{eqnarray}
which, using \eqref{ups:4} and \eqref{ups:9} in \eqref{app:4}, proves \eqref{ups:11} because $T(\infty)=0$.

  Finally, for the integration formulas of $(\UPS')^2$, i.e.,  (\ref{ups:12}--\ref{ups:13}), we refer to Eq.~\eqref{imd:3} in which the relevant calculations are made.

\section*{References}

\end{document}